**Predicting cell-specific gene expression profile and knockout impact through deep learning**

Yongjian He[1,2], Vered Klein[1], Orr Levy[1,#], Xu-Wen Wang[3,#]

[1]*Faculty of Engineering, Bar-Ilan University, Ramat Gan, 52900, Israel*
[2]*International Academic Center of Complex Systems, Beijing Normal University, Zhuhai, 519087, China*
[3]*Channing Division of Network Medicine, Department of Medicine, Brigham and Women's Hospital and Harvard Medical School, Boston, MA 02115, USA*

[#]*These authors jointly supervised this work. Correspondence should be addressed to O.L. (levyorr@biu.ac.il) and X.-W.W. ([spxuw@channing.harvard.edu](spxuw@channing.harvard.edu))*.

# Abstract

Gene expression data is essential for understanding how genes are regulated and interact within biological systems, providing insights into disease pathways and potential therapeutic targets. Gene knockout has proven to be a fundamental technique in molecular biology, allowing the investigation of the function of specific genes in an organism, as well as in specific cell types. However, gene expression patterns are quite heterogeneous in single-cell transcriptional data from a uniform environment, representing different cell states, which produce cell-type and cell-specific gene knockout impacts. A computational method that can predict the single-cell resolution knockout impact is still lacking. Here, we present a data-driven framework for learning the mapping between gene expression profiles derived from gene assemblages, enabling the accurate prediction of perturbed expression profiles following knockout (KO) for any cell, without relying on prior perturbed data. We systematically validated our framework using synthetic data generated from gene regulatory dynamics models, two mouse knockout single-cell datasets, and high-throughput *in vitro* CRISPRi Perturb-seq data. Our results demonstrate that the framework can accurately predict both expression profiles and KO effects at the single-cell level. Our approach provides a generalizable tool for inferring gene function at single-cell resolution, offering new opportunities to study genetic perturbations in contexts where large-scale experimental screens are infeasible.

# Introduction

The gene expression profile is a key determinant of cellular phenotype and function, providing essential insights into the activity and biological significance of the proteins it encodes[1]. Insights from gene expression data help uncover the causes and consequences of diseases, such as classifying different types of cancer, predicting disease recurrence, assessing how various treatments will affect cancer[2,3], the response of cells to disease[4,5], and drug treatments[6,7]. The transcriptional response of a cell to genetic perturbation provides critical insights into cellular function[8,9]. These responses illuminate how gene regulatory networks maintain cell identity and how altering gene expression can reverse disease phenotypes[10–12]. Such understanding is central to biomedical research and the development of personalized therapies[13–15]. For example, using genetic perturbation to validate drug targets has been shown to increase the likelihood of clinical trial success[16,17]. Moreover, identifying synergistic gene pairs can improve the efficacy of combination therapies[18,19].

Although recent advances have accelerated the experimental profiling of genetic perturbations[20,21], the sheer number of possible multigene combinations makes exhaustive testing infeasible[22,23]. Therefore, computational models that can predict perturbation outcomes[24,25], e.g., gene knockout (KO) experiments, are essential for guiding and prioritizing experimental efforts. Several *in silico* KO tools have been developed for this purpose[26,27]. For instance, gene expression-based approaches can predict KO effects[31–33], although some require both wild-type (WT) and KO samples for training[26,27]. Network-based methods, e.g., scTenifoldKnk[31,32] and GenKI[31,32], construct single-cell gene regulatory networks (GRNs) and identify responsive genes by comparing network structures before and after KO perturbations. However, network-based approaches do not estimate cell-type-specific and cell-specific perturbation outcomes[28,33], and other gene expression-based approaches typically require perturbed data and show limited performance[27,34,35]. Many genes remain unexpressed in a cell, and the set of expressed genes in a cell reflects its underlying regulatory state[36], including active transcription factors[37], signaling pathways[38], and chromatin accessibility[39]. This assemblage constrains the expression landscape, enabling accurate prediction of gene expression levels based on the presence or absence of key regulatory and co-expressed genes[40,41].

Here, we propose a data-driven framework to predict the gene expression profile and transcriptional responses to gene KO at single-cell resolution, which does not require the inference of the entire map of gene-to-gene regulatory interactions and perturbed data for model training. Our framework employs a deep-learning model to learn the mapping between gene assemblages and gene expression profiles based only on natural scRNA-seq data, enabling accurate prediction of the transcriptomic response to gene KOs in any cell type and identifying high-impact genes. We systematically validated our framework using simulated gene regulatory networks, where the ground truth of gene interactions and perturbation effects is precisely known. We demonstrated that our data-driven knockout (DKO) framework can accurately predict the cell-specific gene expression profile. The per-cell and cell-type KO impact predicted by DKO is highly correlated with the true impact derived from GRN. We then applied it to analyze two *in vivo* single-cell datasets with KO experiments, finding that those most responsive genes predicted by DKO with a single gene KO in the mouse datasets have also been previously reported in the literature. Finally, we applied DKO to an *in vitro* large-scale CRISPRi-based Perturb-seq dataset from K562 and RPE1 cells[42], encompassing thousands of simultaneous gene perturbations at single-cell resolution. We found that DKO can achieve highly accurate prediction, implying its ability to generalize across perturbations in real biological systems.

# Results

## The DKO framework

Consider a particular cohort of single-cell gene expression data with $N$ different genes, denoted as $\Omega = 1, \cdots, N$. Suppose we have a set of cells $\mathcal{S} = 1, \ldots, M$ collected from a particular cell type of this cohort (**Fig.1a**). Then, we aim to predict the KO outcome of each expressed gene in each cell by using the gene-expression profile only, without knowing or inferring any GRN topology or its dynamics, and without training our model on perturbed data. The DKO framework consists of two phases. In the first phase (**Fig.1b**), we *implicitly learn* the *assembly rules* of genes in a cell type using a deep-learning method with $\mathcal{S}$ as the training data. Since single cells represent distinct steady states, and different cell types may be governed by different gene regulatory networks, this phase captures cell-type–specific regulatory dynamics. This is achieved by learning a map from the gene collection $\boldsymbol{z}$ of a cell $s = (\boldsymbol{z}, \boldsymbol{p})$ to its gene expression profile $\boldsymbol{p}$, i.e., $\varphi: \boldsymbol{z} \mapsto \boldsymbol{p}$. We employed Multi-Layer Perceptron[43] (MLP) to learn such a map, due to its computational efficiency to manage thousands of genes. Here, we utilized MLP to learn the map $\varphi$ for the computational efficacy of over a thousand genes (see **Methods** for details). Learning this map $\varphi$ will enable us to predict the new gene expression profile of a cell upon any gene KO.

In the second phase (**Fig.1c**), to predict the new gene expression profile after a gene-$i$'s KO in a cell $s$, we *conduct a virtual experiment* of knocking out gene-$i$ from the cell $s$ and use the well-trained MLP to compute the impact of gene-$i$'s KO on $s$. In particular, for any cell $s = (\boldsymbol{z}, \boldsymbol{p})$ with the gene collection $\boldsymbol{z}$ and gene expression profile $\boldsymbol{p}$, we remove the gene $i$ from $\boldsymbol{z}$ to form a new gene collection $\tilde{\boldsymbol{z}} = \boldsymbol{z} \backslash i$. Then, for the new gene collection $\tilde{\boldsymbol{z}}$, we use MLP to predict its new expression profile $\tilde{\boldsymbol{p}} = \varphi(\tilde{\boldsymbol{z}})$. To quantify the impact of MLP -$i$'s KO, we compared the new gene expression profile $\tilde{\boldsymbol{p}}$ with the original gene expression profile $\boldsymbol{p}$, and the impact of KO was measured as the dissimilarity $L(\tilde{\boldsymbol{p}}, \boldsymbol{p})$ between $\tilde{\boldsymbol{p}}$ and $\boldsymbol{p}$. In addition, we can summarize the predicted impact scores $L$ of each gene across different cells, i.e., median impact to rank genes, thereby enabling the identification of keystone genes.

## Validation of DKO framework in gene expression profile prediction using synthetic data

To demonstrate DKO's performance in the gene expression profile prediction, we generated synthetic single-cell data using the GRN model with 100 genes and 500 cells (see **Methods**). We simulated single-cell gene expression data using a GRN dynamics model using four parameters: (1) Network topology, e.g., ER random networks[44,45] and scale-free networks[46], which captures the heterogeneity of the number of effector genes regulated by transcriptional factors (TFs). (2) Average degree $\langle k \rangle$. It measures the interaction density among genes, representing the average number of regulatory connections per gene and reflecting the overall complexity of GRN. (3) Difference steady states based on the stochasticity strength parameter $\sigma$, which controls how GRNs differ over cells and the heterogeneity of the single-cell gene expression data accordingly. (4) Dropout rate $d$. To evaluate the DKO in gene expression profile prediction, we randomly split 500 cells into 400 training cells and 100 test cells. DKO was trained to minimize the loss function defined as the mean Bray-Cutis dissimilarity between the true and predicted gene expression for training cells (see **SI Sec.2**). Model performance was then assessed on the test cells by comparing the predicted gene expression profiles with the corresponding ground truth from GRN dynamics.

We first assessed DKO using simulated data generated from various GRN topologies (**Fig.2**). Specifically, we considered three scale-free (SF) networks with power-law exponents $\gamma =$ 2.5, 3.0, and 3.5, as well as an Erdős-Rényi (ER) random network. For each topology, the model was evaluated under a range of parameter settings, including average degree $\langle k \rangle$, stochasticity strength $\sigma$, and dropout rate $d$. Across all network types, DKO consistently achieved high prediction accuracy (**Fig.2a,f,k,p**), with Spearman correlation coefficients ρ around 0.94 (p-value < 1e−4, two-sided t-test) between predicted and true expression levels on the test set, and a higher cell-to-gene ratio can yield more accurate prediction (see **Fig.S1**). The t-SNE projections of gene expression profiles further support these results that predicted data closely match the true expression distribution in two-dimensional space (**Fig.2b,g,l,q**). We also computed the pairwise gene-gene Euclidean distance using true and predicted expression data, respectively. We found that predicted and true distances closely align, while those derived from a randomly perturbed model (e.g., the expression of all genes within each cell is randomly shuffled) diverge significantly (**Fig.2c,h,m,r**). These suggest that our model captured the key gene-to-gene regulatory relationships present in the single-cell expression data.

We further evaluated how predictive performance is affected by three key factors ($\sigma$, $\langle k \rangle$, $d$). Increasing the stochasticity strength $\sigma$ leads to a gradual decrease in prediction accuracy (**Fig.2d,i,n,s**), reflecting the sensitivity of DKO to regulatory heterogeneity across cells. In contrast, changes in the dropout rate $d$ have relatively little effect on performance, suggesting that DKO is robust to moderate levels of missing or sparse gene expression data. Furthermore, model performance improves with a higher average degree $\langle k \rangle$ (**Fig.2e,j,o,t**). This suggests that denser regulatory networks offer a more informative structure for the model to learn from, whereas sparsely connected networks are more challenging and yield lower prediction accuracy. These trends are consistent across all GRN topologies tested, demonstrating the robustness of DKO to both structural and data-level variability.

## Validation of the DKO framework for KO prediction using synthetic data

Accurate prediction of gene expression from a binary gene collection enables us to predict how the outcome will be after the KO of any expressed gene in any cell and measure the impact of KO accordingly. Therefore, we next demonstrated DKO's performance in predicting gene KO using synthetic data. We first trained DKO using all 500 cells, then used well-trained DKO to predict new gene expression profiles associated with each new gene collection $\tilde{z}$ by removing each expressed gene in each cell. The true new gene expression profiles were obtained by integrating the GRN dynamics with the initial condition as the pre-KO gene expression profile and setting the expression level of the target KO gene to 0. Both the true and predicted KO impacts were quantified by the Bray-Curtis dissimilarity between the new post-KO gene expression profile and the original pre-KO gene expression profile.

We systematically assessed DKO using simulated data generated from five various GRN topologies (**Fig.3**). Specifically, we considered scale-free (SF) networks with power-law exponents $\gamma$ = 2.5, 3.0, and 3.5 (each with an average degree of 3), as well as two Erdős-Rényi (ER) random networks with average degrees of 3 and 5. We first compared predicted and true impact scores for all genes across all cells, yielding Spearman correlation coefficients ranging from $\rho$ = 0.915 to 0.97 (**Fig.3a,e,i,m,q**, p-value < 1e-4, two-sided t-test). We also confirmed that this superior performance is not simply attributable to differences in the original pre-KO gene expression profiles (see **Fig.S2**). The distribution of predicted KO impact for the top 10 genes in each network consistently implies that the impact of a gene's KO varies notably across cells, reflecting cell-specific regulatory roles (**Fig. 3b,f,j,n,r**). Moreover, these distributions are shaped by network topology. For instance, in scale-free networks with $\gamma$ = 2.5, a few hub genes exhibit consistently strong effects, whereas in networks with $\gamma$ = 3.5 or in ER networks, where hub structures are less prominent or absent, the impact values become more evenly distributed across genes.

We then summarized the KO impact of each gene across cells into a quantitative impact by its median impact across cells, allowing for the identification of high-impact genes, which reflect their topological characteristics in the GRN (**Fig.S3**). We compared the predicted gene impact rankings to the true rankings (**Fig.3c,g,k,o,s**), observing exceptionally high Spearman correlations between $\rho$ = 0.958 and $\rho$ = 0.981, indicating that DKO captures not only absolute impact but also the relative functional hierarchy of genes. Finally, we framed the identification of high-impact genes as a binary classification task. Using varying thresholds (top 10%-50%) to define positive examples, e.g., high-impact genes, we computed AUROC (Area Under the Receiver Operating Characteristic Curve) values. We found that DKO consistently achieved near-perfect AUROCs (ranging from 0.99 to 1.00), demonstrating its capacity to discriminate between high- and low-impact genes with high accuracy (**Fig.3d,h,l,p,t**). Together, these results demonstrate that DKO not only accurately predicts the magnitude of gene KO effects but also robustly ranks gene importance and identifies key functional regulators across diverse GRN topologies.

## The DKO framework detects knockout-induced transcriptional outcomes in in vivo mouse single-cell data

Next, we tested our DKO framework using two *in vivo* single-cell cohorts with gene knockouts: (1) a study investigating the transcriptional control of lung alveolar type 1 cell development and maintenance by NK homeobox 2-1 (*Nkx2-1*) using mouse models[47], and (2) a study demonstrated that chronic phagocytic stress due to demyelination leads to a similarly reduced expression of lysosomal and lipid metabolism genes in microglia isolated from the brains of *Trem2* knockout mice[48]. Each dataset includes both control (pre-KO) and perturbed data (post-KO). In each dataset, we focused on the two cell types with the largest number of cells (see **Methods** for details). In the *Nkx2-1* KO cohorts, Plasma cells (3,214 genes; 2,739 cells) and plasmacytoid dendritic cells (pDC; 3,214 genes; 1,425 cells) were retained. In the *Trem2* KO cohorts, the selected cell types were CX3CR1$^+$ macrophages (3,214 genes; 740 cells) and SPP1$^+$ macrophages (4,895 genes; 556 cells).

To evaluate the DKO in predicting gene expression profiles in these four cellular cohorts, we randomly split 80% of cells from pre-KO mice to train DKO and the remaining 20% of cells as the test set for each cohort. We found that DKO can accurately predict gene expression across four cohorts (see **Fig.4a,e,i,m**). The Spearman correlation $\rho$ between the true expression level and the predicted expression level is 0.97~0.99 with a p-value<1e-4 (two-sided t-test). To examine whether violating the underlying hypothesis, e.g., the mappings between gene assemblages and expression in different cell types are governed by distinct GRNs, reduces the performance of DKO, we also trained DKO using cells from all cell types, e.g., without separating into different cell types. We found that DKO showed a limited performance, suggesting that different cell types were governed by distinct GRNs, which also explains the results that generic deep learning approaches did not yield superior performance than naïve baseline models[49,50].

We then applied the DKO framework to the four cohorts to predict the KO outcome on the cellular gene expression, respectively. We first trained DKO using all pre-KO cells for each cohort and then applied it to predict post-KO gene expression profiles. For each gene (excluding the KO target), we summarized the knockout effect as a gene-level change defined as the mean expression across post-KO cells minus the mean across pre-KO cells. We computed this quantity twice, e.g., using the observed post-KO data (true change) and using the DKO-predicted post-KO profiles with the same pre-KO baseline (predicted change), respectively. We find the predicted and true gene-level changes show strong concordance (**Fig.4b,f,j,n**), with the Spearman correlation $\rho$ being 0.57~0.91(p-value<1e-4, two-sided t-test).

To evaluate DKO's ability to prioritize the most affected genes, we computed top-k overlap between predicted and true rankings, where genes were ranked by the absolute magnitude of their effects. The observed overlap consistently exceeded the random baseline $k/N$ across a wide range of $k$ (**Fig.4c,g,k,o**), indicating that DKO preferentially elevates truly perturbed genes.

At the single-cell level, we assessed variability using a pseudo-bulk procedure. For each replicate, we randomly sampled 30 cells from the KO dataset and 30 cells from the corresponding pre-KO dataset and computed the mean gene-profile change (KO minus pre-KO). Repeating this 100 times yielded distributions of KO-induced changes. For the top 10 genes ranked by true effects, predicted and observed distributions closely matched (**Fig.4d,h,l,p**), indicating that our DKO model captures KO-driven shifts not only in aggregate but also under repeated cell-level subsampling.

Next, we tested the DKO framework's ability to identify the genes that were most highly impacted (e.g., expression level difference before and after KO) by KO of the *Nkx2-1* gene and *Trem2,* respectively. **Fig.4d,h,l,p** shows the most influenced 20 genes whose expression levels were increased after KO (yellow) or decreased (blue) in four cell types. We found that the impact of *Nkx2-1* KO displays a large variation for those 10 downregulated genes, while its impact for those upregulated genes is quite stable over different cells. We found that the genes most influenced are highly correlated with *Nkx2-1*. For example, DKO predicted that the KO of *Nkx2-1* would decrease the expression level of *SFTPB* and *SFTPC*, which was consistent with previous findings that transfection of A549 cells with *Nkx2-1* expression vectors demonstrated decreased stimulation of *Sftpb* and *Sftpc* expression by mutant proteins compared with that of WT in infant[51]. *Cxcl15* is a marker gene of alveolar type II (ATII) identity that can be driven by *Nkx2-1* in lung cancer[52]. Interestingly, among those top 10 downregulated genes, 5 genes, *CD74*, *Naspa*, *Wfdc2*, *Sftpc*, and *Ager* were also found as KO-responsive genes identified by GenKI[32], while none of the top 10 upregulated genes overlapped. Conversely, overexpression of NKX2-5 suppressed the expression of contractile genes (*ACTA2, TAGLN, CNN1*). *Trem2* potentially supports the activation of microglia initiated by other receptors in stage 1 of disease-associated microglia (DAM), characterized by the expression of genes such as *Cst3* and *Hexb*[53,54]. *Ctss* is also a DAM gene related to the *Trem2* gene dosage reprograms[55]. The expression of *Tmsb4x, C1qa, C1qb* was considered a feature of homeostatic microglia[56].

## Validation of DKO framework using Perturb-seq data

Finally, we evaluated the performance of the DKO framework using a publicly available genome-scale Perturb-seq dataset (see **Methods** for details) based on CRISPR interference (CRISPRi)[42], including samples from two cell lines, K562 and RPE1 (**Fig.5**). Each cell line consists of a control group (non-perturbed cells with non-targeting sgRNAs) and a perturbation group (cells perturbed by gene-specific sgRNAs). As illustrated in **Fig.5a**, we used control cells to train and validate the DKO model for gene expression prediction, with 80% of control cells used for training and the remaining 20% reserved for testing. For KO outcome prediction, the DKO model was trained using control groups and was applied to the perturbation group for evaluation. For each targeted gene, we constructed pseudo-bulk profiles by aggregating the transcriptomes of cells carrying the corresponding sgRNAs. The model predicted expression profiles from gene assemblages, and KO impact was quantified as the Bray-Curtis dissimilarity between the predicted pseudo-bulk and a baseline pseudo-bulk derived from control cells.

We found that, in both cell lines, DKO achieved high accuracy in predicting gene expression levels (**Fig.5b** for K562 cell line, **Fig.5g** for RPE1 cell line), with Spearman correlation coefficients of $\rho = 0.974$ for K562 and $\rho = 0.980$ for RPE1 (both p-value $< 1e-4$, two-sided t-test). For KO outcome prediction, ground truth KO impact for each gene was calculated by aggregating the transcriptomes of all perturbed cells into a pseudo-bulk profile and comparing it against a control pseudo-bulk (constructed from unperturbed cells) using Bray-Curtis dissimilarity. We found that the predicted impact scores closely matched the ground truth values, yielding Spearman correlations of $\rho = 0.925$ for K562 (**Fig.5c**) and $\rho = 0.954$ for RPE1 (**Fig.5h**, p-value $< 1e-4$, two-sided t-test). The slightly lower performance on K562 is due to the model's tendency to underestimate KO impact for genes with both high cross-cell mean gene count and high cross-cell standard deviation of that count (see **Fig.S4**).

We also assessed DKO's ability to identify high-impact genes by framing the task as a binary classification problem (**Fig.5d,i**). Genes were ranked by their ground truth KO impact, and top-k thresholds (e.g., top 10%, 20%, 30%, etc.) were used to define positive labels. The resulting ROC curves and AUC scores confirm excellent discriminative performance, with AUCs ranging from 0.930 to 0.965 in K562 and from 0.946 to 0.988 in RPE1. To test our hypothesis that cell types are governed by distinct GRNs, we trained the model on the RPE1 cell line and evaluated DKOs in K562, observing a low correlation between true and predicted gene ranks (Spearman ρ = 0.15) and an AUC of 0.56 for identifying high-impact genes (see **Fig.S5**). Finally, we also presented the distribution of predicted KO impact for the top 10 genes (**Fig.5e,j**) and the bottom 10 genes (**Fig.5f,k**) in each cell type, ranked by their median KO impact across cells. We identify *SUPT6H* (SPT6) and *SUPT5H* (SPT5) as top-impact genes in K562, and prior work shows SPT6 couples RNA polymerase II elongation to nucleosome reassembly[57], while SPT5 regulates the pause near promoters and promotes release into productive elongation[58]. In RPE, top-impact genes SMN2 and TRNT1 are supported by studies linking SMN deficiency causes retinal abnormalities[59], hypomorphic TRNT1 mutations lead to retinitis pigmentosa[60]. Overall, these results highlight the robustness and generalizability of DKO in real-world, high-throughput functional genomics applications.

## Discussion

The knockout impact of a gene could be cell-specific due to the complex interactions, including both direct and indirect interactions between the target knockout gene and its neighbors. Despite many computational tools that have been developed to predict the knockout impact, for instance, the responsive genes, accurately estimating cell-specific knockout outcomes remains highly challenging. This difficulty arises from our limited understanding of gene regulatory network dynamics in individual cells, as well as the logistical and ethical constraints of experimentally perturbing thousands of genes. In this work, we took a novel approach, that does not require the extraction of the entire map of gene-to-gene regulatory interactions and propose a data-driven framework to systematically predict the KO impact of each expressed gene in each single cell. Our framework can be used to facilitate data-driven dissecting of the roles of specific genes, understanding disease mechanisms, and developing new gene therapies.

In our current framework, we used MLP to learn the map $\varphi: \boldsymbol{z} \mapsto \boldsymbol{p}$. However, we emphasize that the proposed framework is general enough and can be modified in many different ways. For example, instead of using MLP, one can use other deep learning models, e.g., ResNet and neural ODE as long as the training of thousands of genes is feasible for the deep learning model and the sample size (number of cells) is approximately twice the number of genes. In addition, instead of using Bray-Curtis dissimilarity, one can use other dissimilarity or distance measures, e.g., the weighted Euclidean distance, to quantify the impact of genes' KO. Unlike using the definition of keystones, which combines the impact component and the biomass component to quantify the removal of a species from a microbial community, we did not consider the contribution of the gene's expression level. Moreover, our framework can be naturally extended to predict the impact of many genes' KO.

# Methods

**Synthetic single-cell data.** We generated synthetic single-cell data using a general GRN model[61] based on Michaelis–Menten dynamics. The gene expression of each simulated cell was chosen to be the steady state of a coupled ordinary differential equation (ODE), which represent the dynamics of a group of genes with regulatory interactions. The gene expression state was randomly initialized such that 80% of the genes in each cell were expressed, with their initial expression levels sampled from a uniform distribution $\mathbb{U}(0.1,1)$, while the remaining 20% were unexpressed, i.e., their expression values were set to 0 at initialization and remained 0 throughout the dynamics. The ODE below describes the dynamics of gene $i$ expression levels in the model:

$$\frac{dx_i}{dt} = -B_i x_i + \sum_j w_{i,j}^{act} \frac{x_j}{1+x_j} + \sum_j w_{i,j}^{rep} \frac{1}{1+x_j},$$

$B_i$ is the degradation rate of gene $i$ (we set the degradation to be 1 for all genes). The activation and repression terms describe nonlinear regulatory inputs from other genes through weighted interactions.

We varied four parameters to generate diverse GRN-based single-cell data: (1) Network topology. We considered Erdős-Rényi (ER) random networks and scale-free (SF) networks. These two topologies allowed us to investigate different regulatory architectures within the GRN framework. In both cases, nodes were defined as transcription factors (TFs) or effector genes, with equal numbers of each. Regulatory interactions were restricted to TF-TF and TF-effector edges. (2) Average degree $\langle k \rangle$. We considered different average degrees of network to capture varying levels of regulatory complexity. (3) Stochasticity strength parameter $\sigma$. All cells in a cohort or cell type shared the same GRN topology and a base weighted regulatory matrix $W^*$, with nonzero entries ($w_{(i,j)}^* > 0$) sampled from uniform distribution $\mathbb{U}(0,2)$. For each cell $v$, its weighted regulatory matrix $W^v$ was generated by perturbing $W^*$: each nonzero entry was retained with probability $1 - \sigma$, or replaced by a new draw from uniform distribution $\mathbb{U}(0,2)$ with probability $\sigma$. This introduced cell-to-cell heterogeneity while preserving the underlying network. (4) Dropout rate $d$. After solving the GRN ODEs to steady state, dropout was applied to mimic single-cell RNA-seq sequencing depth. When $d = 0$, values were unchanged. For $d > 0$, genes below the 20th percentile were set to zero with probability $d$ (for $d > 0.1$), otherwise a baseline dropout rate of 0.1 was used.

**Real single-cell data.** We used two real mouse single-cell datasets: NK homeobox 2-1 (*Nkx2-1*) mouse lung cells [47] and *Trem2* knockout mouse brain microglia[48]. The single-cell gene expression data was processed by Seurat[62,63]. To normalize the data, we applied Seurat's NormalizeData function using the "LogNormalize" method and a scaling factor of 10,000, which transforms the gene expression values into log-transformed, normalized counts. Subsequently, genes with low prevalence across cells were removed. Specifically, genes that were expressed in less than 20% of the cells (i.e., with counts greater than 0 in less than 20% of the columns) were filtered out.

After obtaining the filtered single-cell data, we performed cell type annotation using CellTypist, an open-source Python package for machine learning-based classification of single-cell transcriptomes with pretrained reference models[64,65]. Candidate pretrained models were selected according to the extent of overlap between their feature genes and those present in the filtered single-cell dataset. Cell type labels were then assigned using CellTypist's majority voting strategy, which consolidates predictions across decision paths to produce robust cell type assignments.

Specifically, *Nkx2-1 KO* dataset was annotated using the *Adult_Mouse_Gut* model, yielding two major cell cohorts: Plasma cells (3,214 genes; 2,739 cells) and plasmacytoid dendritic cells (pDC) (3,214 genes; 1,425 cells). *Trem2* KO dataset was annotated using the *Cells_Fetal_Lung* model, yielding two dominant cell cohorts CX3CR1$^+$ macrophages (3,214 genes; 740 cells) and SPP1$^+$ macrophages (4,895 genes; 556 cells).

From these annotations, we ultimately obtained four single-cell cohorts, which we used to validate our DKO experiments.

**Real Perturb-seq data.** We also used a single-cell Perturb-seq dataset[42], which integrates a compact, multiplexed CRISPR interference (CRISPRi) library with single-cell RNA sequencing (scRNA-seq) to profile thousands of loss-of-function perturbations at single-cell resolution. The Perturb-seq design encompassed two scales of genetic perturbations: a genome-wide set covering 9,867 genes and an essential-wide subset comprising 2,285 common essential genes.

This dataset was generated in two human cell types: chronic myeloid leukemia (CML, K562) and retinal pigment epithelial (RPE1). For each cell type, two categories of single-cell data were available: non-targeting perturbations (unperturbed control cells) and targeting perturbations (perturbed cells). Specifically, for the K562 cell type, we used the essential-wide subset, containing 10,691 control cells and 299,694 perturbed cells, whereas for the RPE1 dataset, we used the genome-wide set, which contains 11,485 control cells and 236,429 perturbed cells.

In both cases, we used the single-cell control data to train our model, and the single-cell perturbed data to evaluate the performance of the DKO model. The availability of experimentally perturbed transcriptomes in the dataset provides ground-truth references, allowing us to directly assess the accuracy of our model in predicting gene expression profiles and KO outcomes.

**Deep learning model.** To learn the mapping from the expressed gene assemblage to expression levels, we first transform the gene collection $\mathbf{z} \in \{0,1\}^N$ into the normalized collection $\tilde{\mathbf{z}} = \mathbf{z}/\mathbf{1}^T\mathbf{z} \in \Delta^N$, where $\mathbf{1} = (1,\cdots,1)^T$. Then, the gene assemblage $\mathbf{z}$ was fed to a two-hidden layer MLP model:

$$f_w(\mathbf{z}) = w_2(w_1(\mathbf{z}))$$

$w_1 \in \mathbb{R}^{(N_g \times N_g)}$, $w_2 \in \mathbb{R}^{N_g \times N_g}$ to learn more complex patterns in the data. $N_g$ represent the number of genes. Finally, an element-wise multiplication and normalization were performed:

$$\mathbf{p} = \mathbf{z} \odot f_w(\mathbf{z}) / \sum \mathbf{z} \odot f_w(\mathbf{z})$$

This will guarantee the sum expression of all genes is 1 and the expression level of those genes not expressed is always 0.

MLP is trained to minimize the loss function:

$$E(\mathcal{D}) = \frac{1}{|\mathcal{D}|} \sum_{((\mathbf{z},\mathbf{p})\in\mathcal{D})} d(\mathbf{p}, \hat{\varphi}_w(\mathbf{z})).$$

Here $\mathcal{D}$ is the data set, and $d$ can be any distance or dissimilarity measure, e.g., the Bray-Curtis dissimilarity. We used different batch sizes for different datasets due to the sample size. The batch size is 20, and the total epoch is 1,000 for all datasets.

**Acknowledgements.** X.-W.W. acknowledges the funding support from National Institutes of Health (K25HL166208).

**Author Contributions.** X.-W.W. conceived the project. O.L. and X.-W.W. designed the project. Y.-J.H. performed all the numerical calculations and real data analyses. Y.-J.H., O.L., and X.-W.W. wrote the manuscript. All authors contributed to the interpretation of the results and revised the manuscript.

**Competing Interests Statements.** The authors declare no competing interests. Correspondence and requests for materials should be addressed to O.L. (levyorr@biu.ac.il) and X.-W.W. (spxuw@channing.harvard.edu).

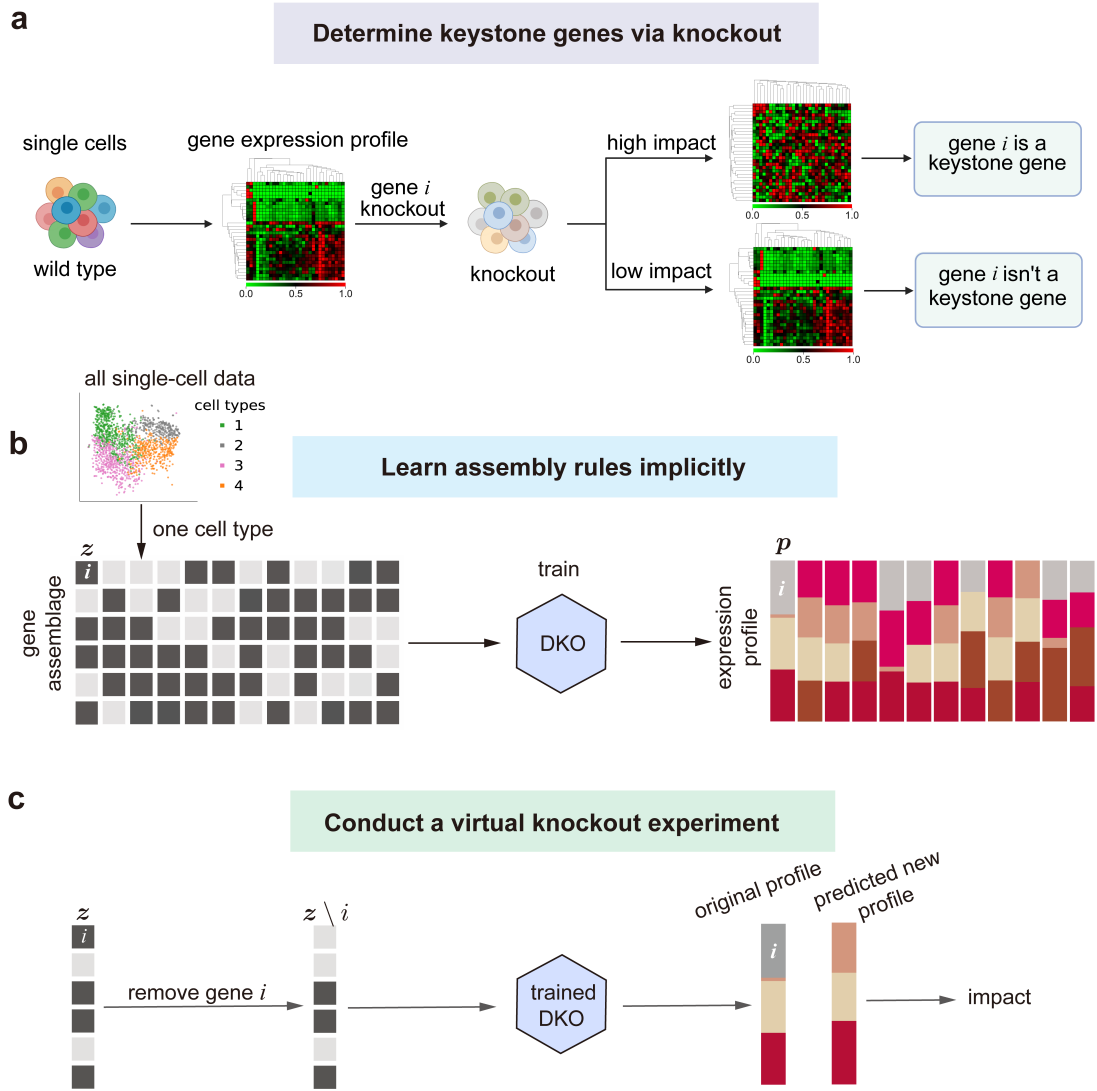

**Figure 1: Workflow of the data-driven knockout (KO) prediction (DKO) framework. a,** Given the gene expression profiles of a set of cells from a particular cell type, we determine if a gene is a high-impact gene (keystone gene) by comparing the impact of its knockout, e.g., the dissimilarity between new expression profiles after its knockout versus original expression profiles before knockout across all cells. **b,** The gene assemblage/collection of a single-cell sample $s$ is represented by a binary vector $\boldsymbol{z} \in \{0,1\}^N$, where its $i$-th entry satisfies $z_i = 1$ ($z_i = 0$) if gene-$i$ is expressed (or not expressed) in this cell. The gene expression profile of the cell is characterized by a vector $\boldsymbol{p}$, where its $i$-th entry $q_i$ is the expression level of gene-$i$. A deep learning model (MLP) is trained to learn the map: $\boldsymbol{z} \in \{0,1\}^N \mapsto \boldsymbol{p} \in \Delta^N$ using the training cells. **c,** We conduct a *thought KO* by removing a gene from the gene assemblage to obtain a new assemblage. Then, for the new gene assemblage, we use MLP to predict its new gene expression profile $\tilde{\boldsymbol{p}} = \varphi(\boldsymbol{z})$. The KO impact of this gene in this cell is measured by the distance between the original gene expression profile $\boldsymbol{p}$ and new gene expression profile $\tilde{\boldsymbol{p}}$.

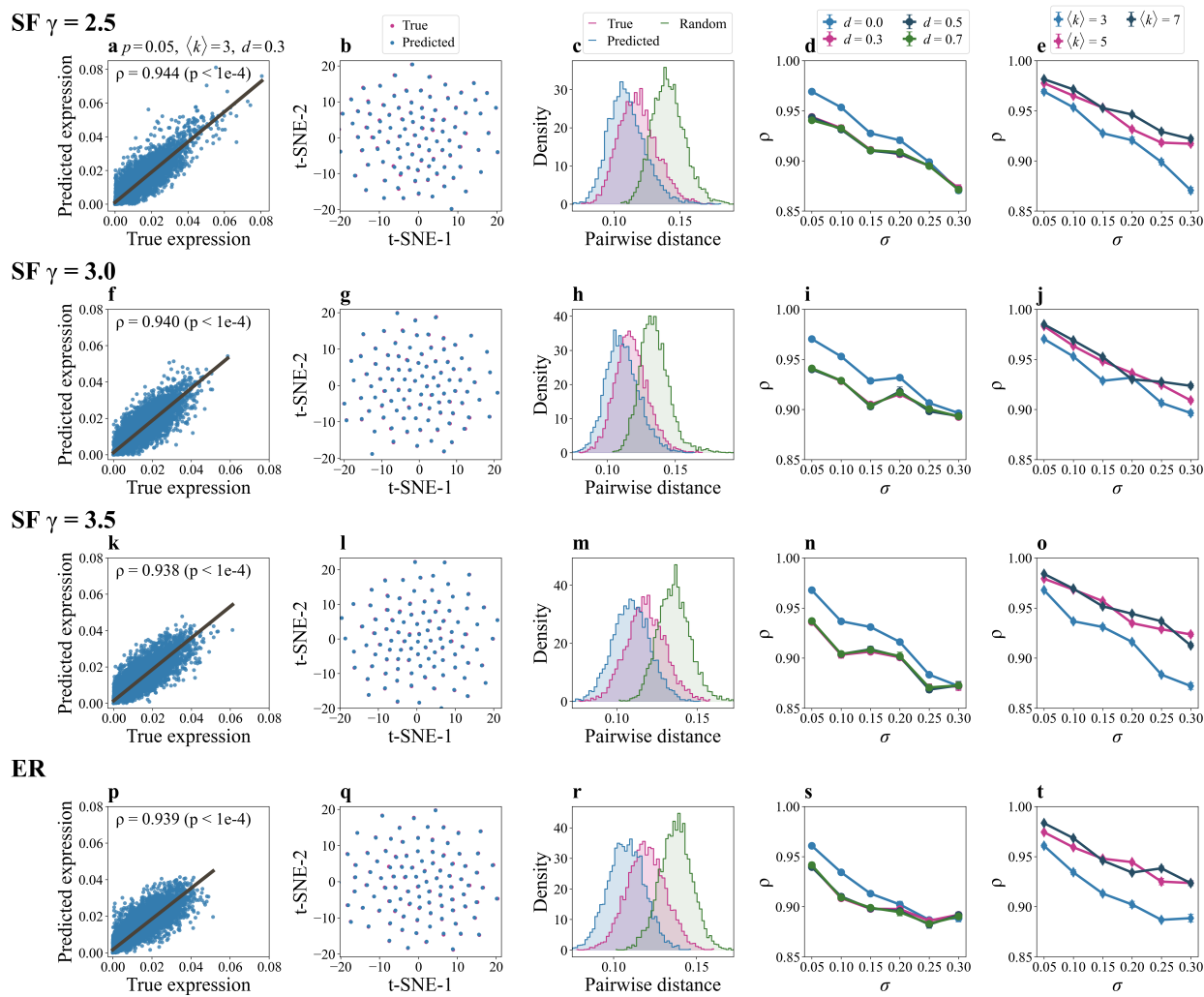

**Figure 2:** *In silico* **validation of the DKO framework in gene expression prediction.** Results are obtained for the pools of $N = 100$ genes under four gene regulatory network (GRN) topologies: scale-free (SF) networks with power-law exponents $\gamma = 2.5$, 3.0, and 3.5, and Erdős-Rényi (ER) random networks. For each setting, we used 400 cells to train DKO and the remaining 100 to validate the performance of DKO. **(a, f, k, p)** Scatter plots comparing predicted and true gene expression values in test cells. Each panel reports the Spearman correlation coefficient ρ and the corresponding p-value (computed using a two-sided t-test). **(b, g, l, q)** t-SNE projections of gene expression profiles, where each point represents a gene embedded from the original 100-dimensional expression space (with each dimension corresponding to one test cell). **(c, h, m, r)** Distributions of pairwise Euclidean distances between genes, computed based on their expression vectors across the 100 test cells. **(d, i, n, s)** Prediction performance under varying stochasticity strength $\sigma$ and dropout rate $d$. **(e, j, o, t)** Prediction performance under varying stochasticity strength $\sigma$ and average degree $\langle k \rangle$.

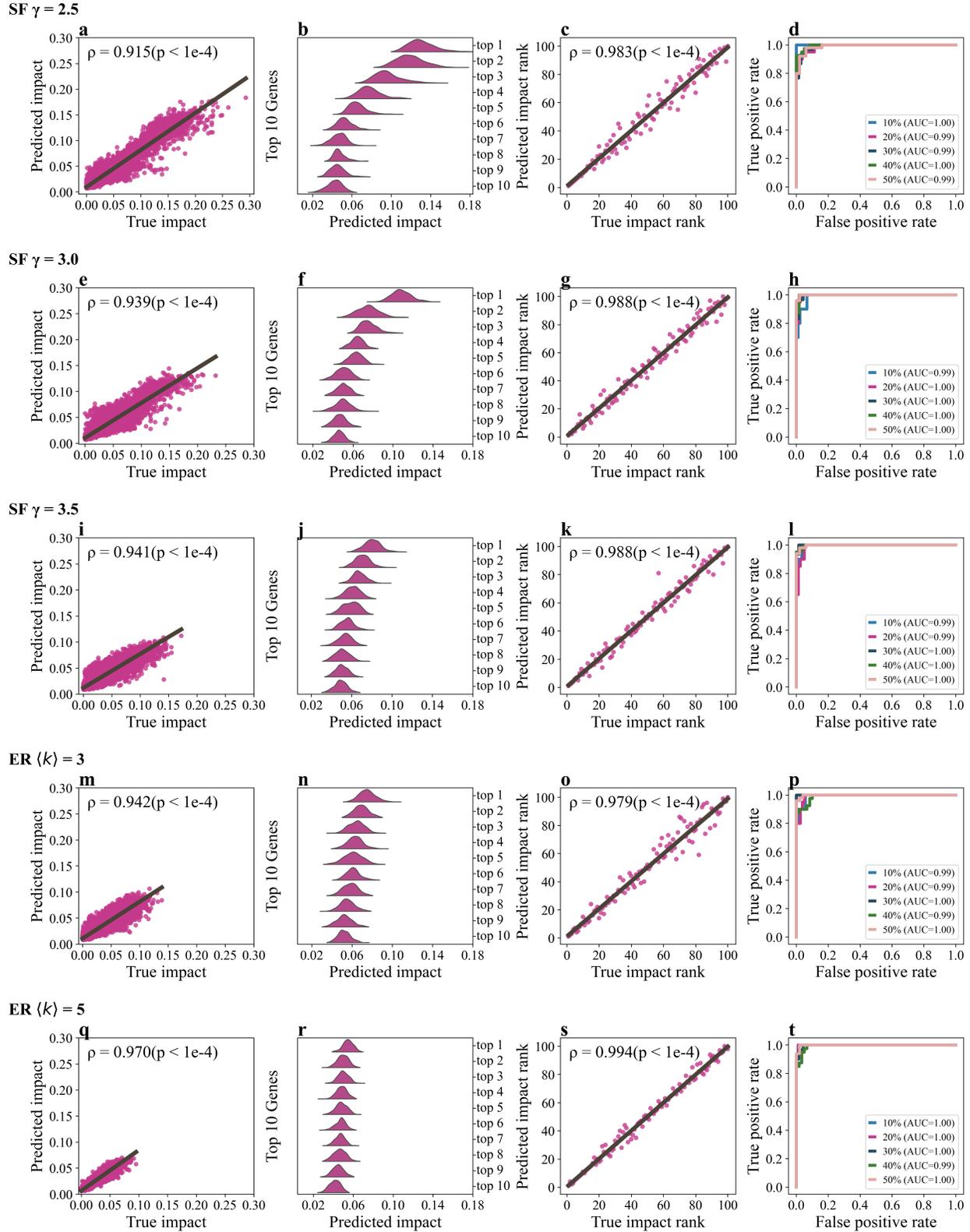

**Figure 3: In silico validation of the DKO framework in KO prediction.** Results are obtained for the pools of $N = 100$ genes under five gene regulatory network (GRN) topologies: three scale-free (SF) networks with power-law exponents γ = 2.5, 3.0, and 3.5 (all with an average degree of 3), and two Erdős-Rényi (ER) random network with average degrees of 3 and 5. For each network, the stochasticity strength $\sigma$ was set to 0.15, and dropout rate $d$ set to 0. We used 500 cells to validate the performance of DKO. **(a, e, i, m, q)** Predicted KO impact versus true KO impact values, quantified by Bray-Curtis dissimilarity between expression profiles before and after KO. Each panel also shows the Spearman correlation coefficient $\rho$ and the corresponding p-value (two-sided t-test). **(b, f, j, n, r)** Distributions of predicted KO impact scores for the top 10 genes ranked by median predicted impact across cells. **(c, g, k, o, s)** Predicted rankings versus true rankings of gene impact scores. **(d, h, l, p, t)** Receiver operating characteristic (ROC) curves evaluating the ability of DKO to prioritize high-impact genes. Genes with top KO impact (e.g., top 10%, 20%, 30%, etc., based on ground truth) were treated as positive instances, and AUC values were used to assess how well the predicted impact scores ranked these genes ahead of others.

**CX3CR1**

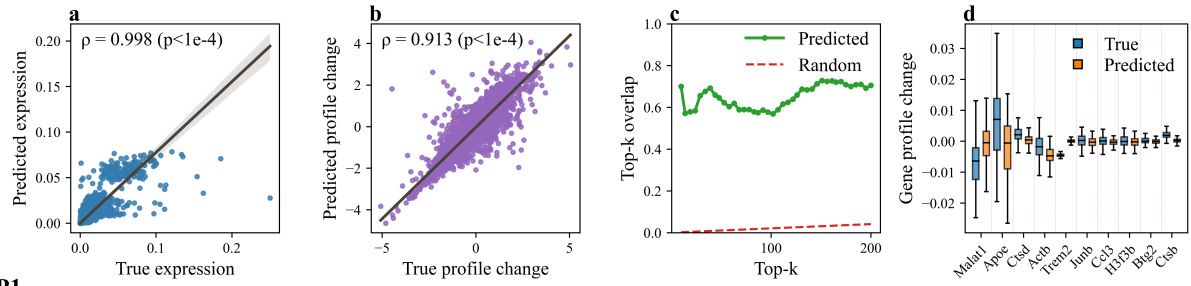

**SPP1**

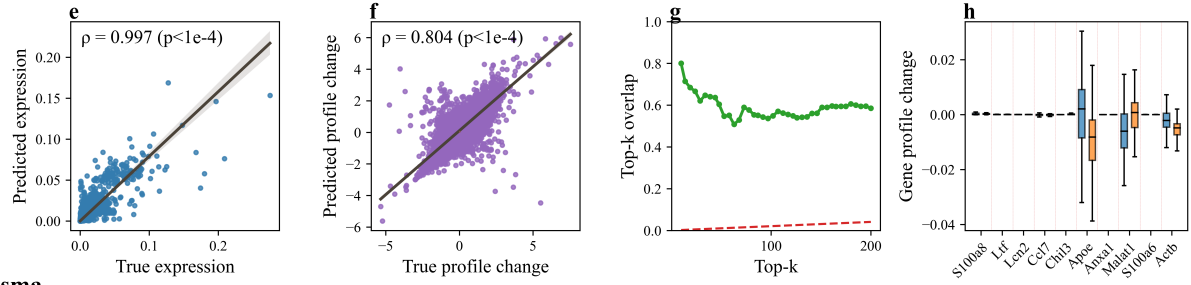

**Plasma**

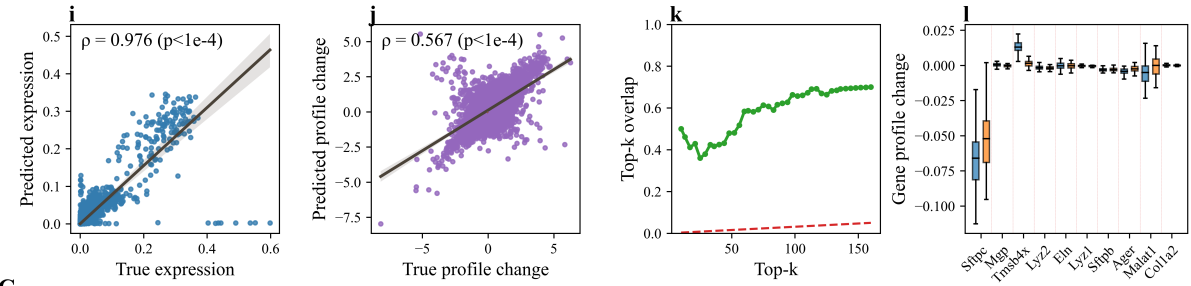

**pDC**

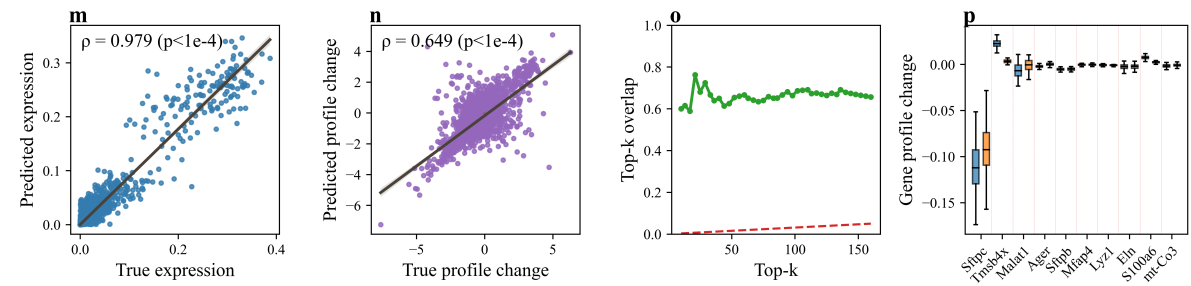

**Figure 4: Validation of the DKO framework in gene expression and KO outcome using real single-cell data.** (**a,e,i,m**) Validation of gene expression prediction. For each cell cohort (CX3CR1 and SPP1 in *Trem2 KO* dataset and plasma and pDC in *Nkx2-1, KO* dataset), we randomly split 80% of pre-KO cells to train DKO and the remaining 20% of pre-KO cells as the test set. Each dot represents the true and predicted relative gene expression level of each gene in each of the test cell. Each panel also displays the Spearman correlation coefficient ρ and the corresponding p-value (a two-sided t-test). (**b,f,j,n**) Validation of KO outcome. We trained the model using all pre-KO data and then used it to predict the gene expression profiles after KO. Each dot represents the predicted versus true profile change of each gene. For each gene, the predicted change was calculated as the mean profile across all cells in the predicted post-KO data minus the mean profile in the corresponding pre-KO data. The true change was calculated analogously, using

the true post-KO data mean minus the pre-KO mean. (**c,g,k,o**) Top-k overlap between the predicted and true KO-induced gene profile changes, with random overlap (k/N) as baseline (N is the number of genes). Overlap was calculated based on the top-k genes ranked by the absolute magnitude of predicted and true effects. (**d,h,l,p**) Cell-level predicted and true KO-induced changes for the top 10 most affected genes. For each cell type, we applied a pseudo-bulk strategy to generate multiple replicates. In each replicate, 30 cells were randomly sampled from the true or predicted KO data and 30 cells from the corresponding pre-KO data. The mean gene profile change (KO minus pre-KO) was then computed. Repeating this procedure 100 times yielded distributions of KO-induced changes across cells, which are visualized as boxplots for true (blue) and predicted (orange) data.

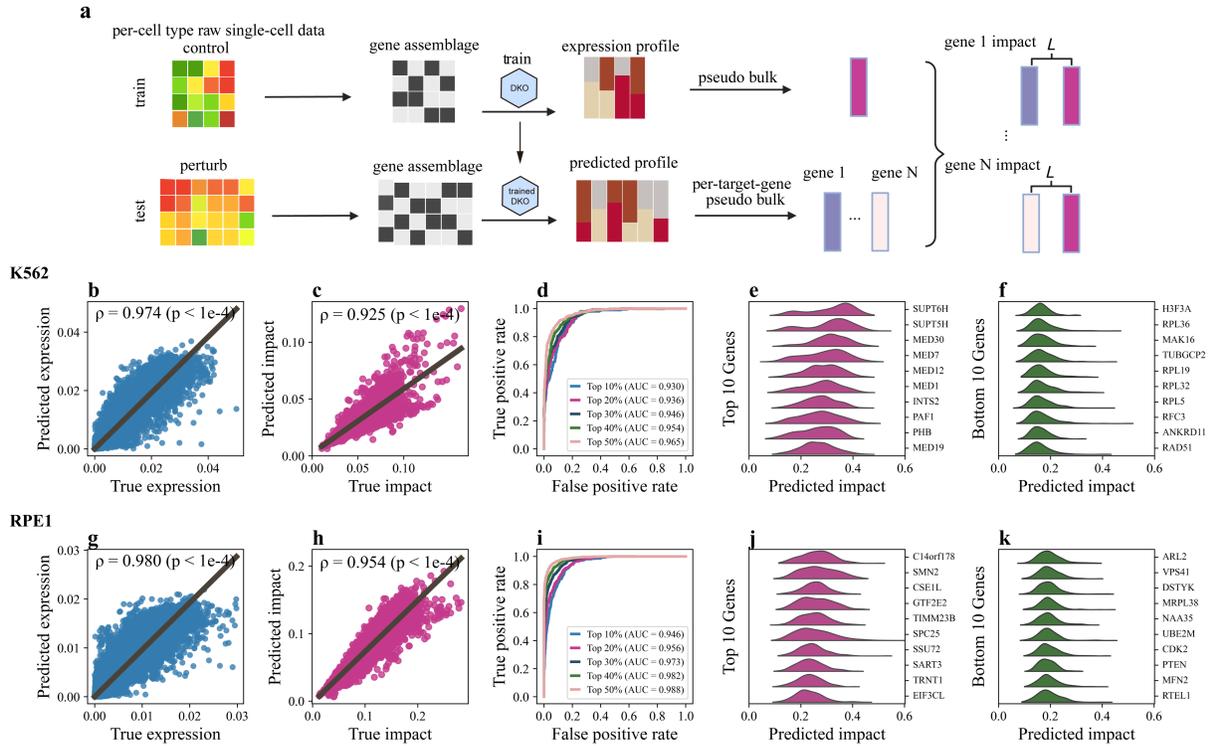

**Figure 5: Validation of the DKO framework on CRISPRi Perturb-seq datasets of two cell lines.** We evaluated DKO on a large-scale CRISPR interference (CRISPRi) Perturb-seq datasets containing two distinct cell types, derived from K562 and RPE1 cells. **(a)** Schematic of the evaluation workflow. Each dataset was split into a control group (non-targeting sgRNAs) and a perturbation group (targeted gene knockdowns). The control group was used to train and test the model for gene expression prediction (80% training, 20% testing). KO impact was quantified by computing the Bray-Curtis dissimilarity between pseudo-bulk transcriptomes. The true KO impact was computed between the observed pseudo-bulk of perturbed cells and the pseudo-bulk of control cells. The predicted KO impact was computed between the DKO-predicted pseudo-bulk and the same control pseudo-bulk. **(b-d)** Results on the K562 cell type. **(e-g)** Results on the RPE1 cell type. **(b, g)** Gene expression prediction in control cells. Each panel includes the Spearman correlation coefficient ρ and associated p-value (computed using a two-sided t-test). **(c, h)** Predicted KO impact versus true KO impact for each gene. **(d, i)** ROC curves and AUC values evaluating DKO's ability to identify high-impact genes, using top-k KO thresholds (e.g., top 10%, 20%, etc.) based on ground-truth impact. (**e**,**f**) Predicted KO impact distributions across single cells for the top 10 **(e)** and bottom 10 **(f)** genes in the K562 cell type. **(g-k)** Predicted KO impact distributions across single cells for the top 10 **(g)** and bottom 10 **(k)** genes in the RPE1 cell type. The top and bottom genes were selected based on the median predicted KO impact across cells for each gene.

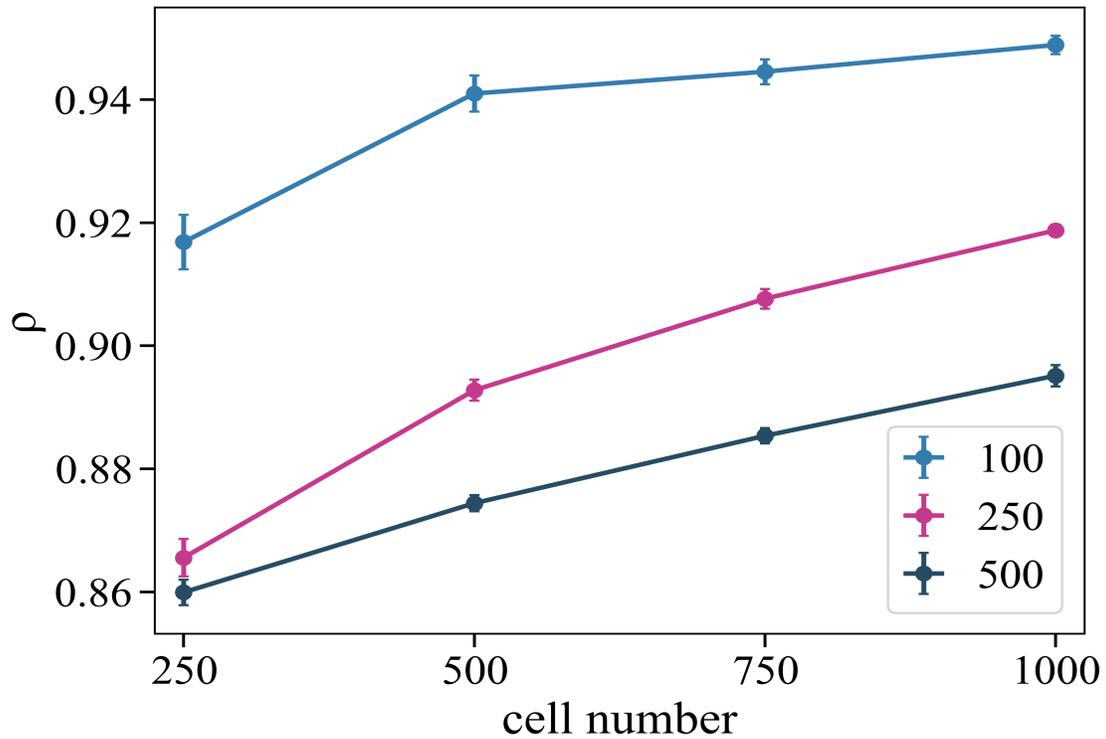

**Figure S1**: **Gene expression profile prediction varies with the different number of genes and cells in simulated data from a scale-free GRN ($\gamma$ = 2.5).** Each configuration shows the mean Spearman correlation $\rho$ over 10 independent runs. Performance is high overall ($\rho \geq 0.86$) and increases with the number of cells.

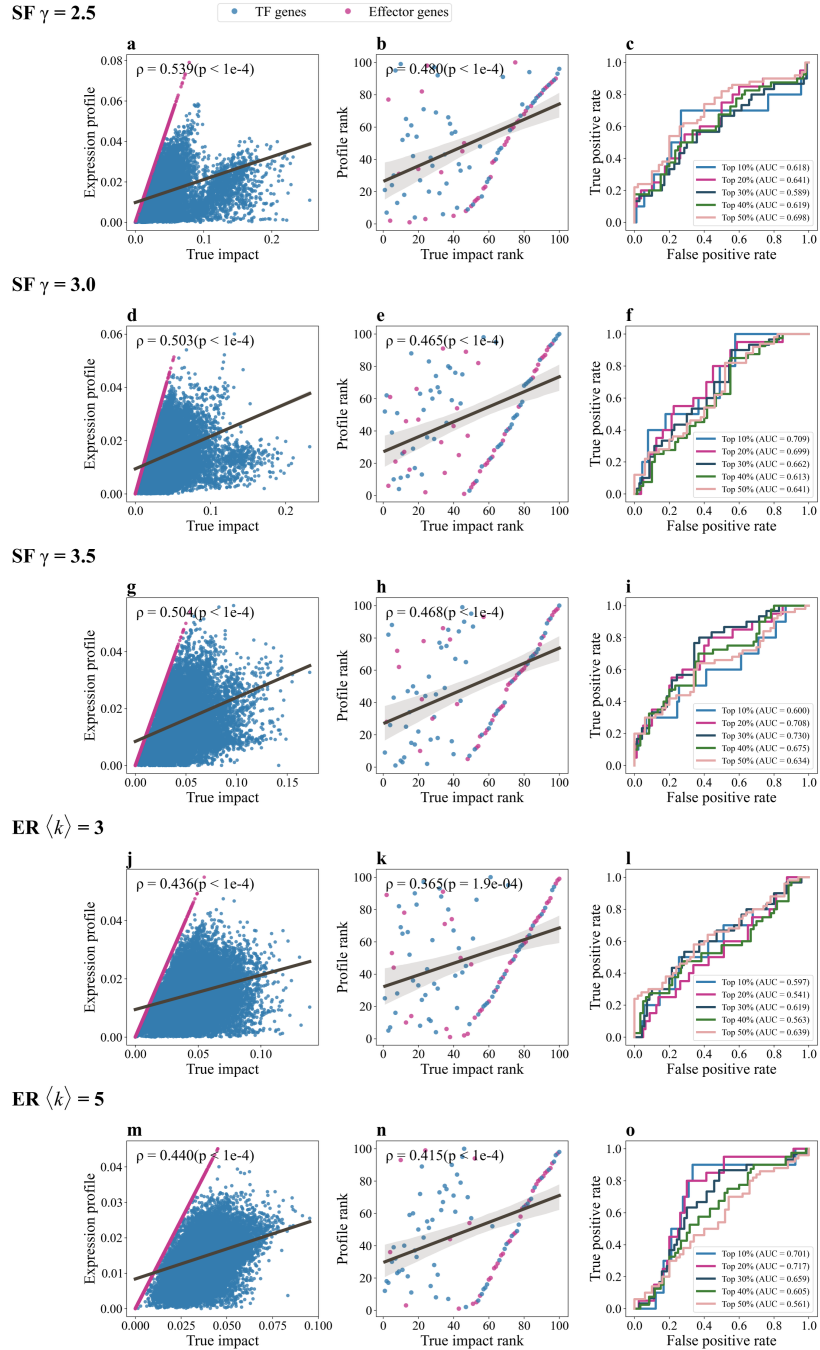

**Figure S2: Correlation between ground truth gene KO impact and gene expression level.** **(a,d,j,g,m)** Ground-truth gene KO impact versus gene expression for transcriptional factors (blue) and target genes (purple). **(b,e,h,k,n)** Spearman rank correlation between ture impact rank and expression rank. **(c,f,I,l,o)** AUROC when casting identification of key genes as a binary classification using expression alone. Overall, associations are weak and AUROCs modest, indicating that gene importance is driven primarily by regulatory connectivity rather than expression level. Notably, some transcription factors show low expression yet regulate many targets and therefore exhibit high KO impact.

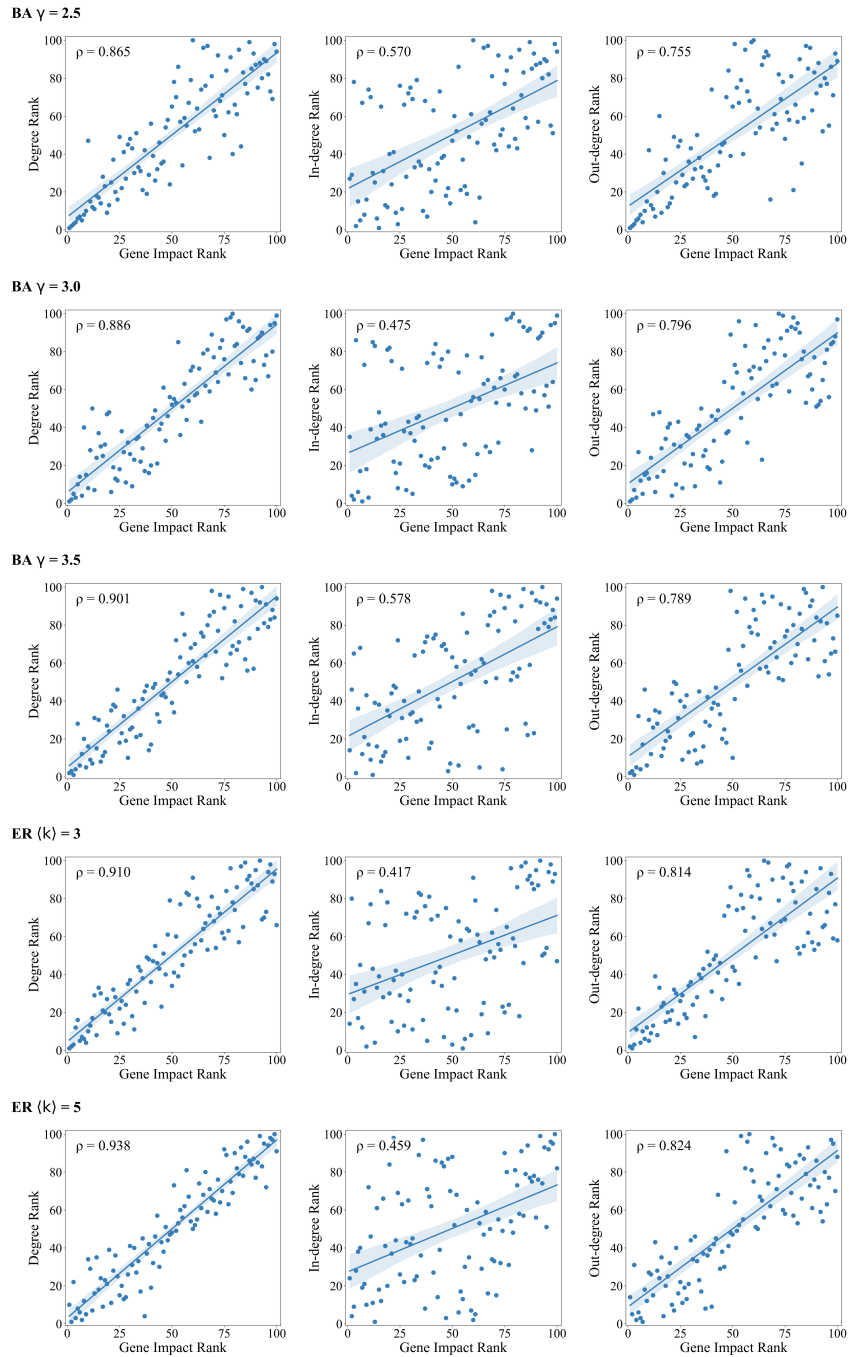

**Figure S3: Correlation between true gene KO impact and gene network features.** For each gene, in-degree is the number of upstream regulators and out-degree is the number of downstream targets, and the total degree is the sum of the in- and out-degrees. Overall, the rank correlation with total degree was the strongest, followed by out-degree.

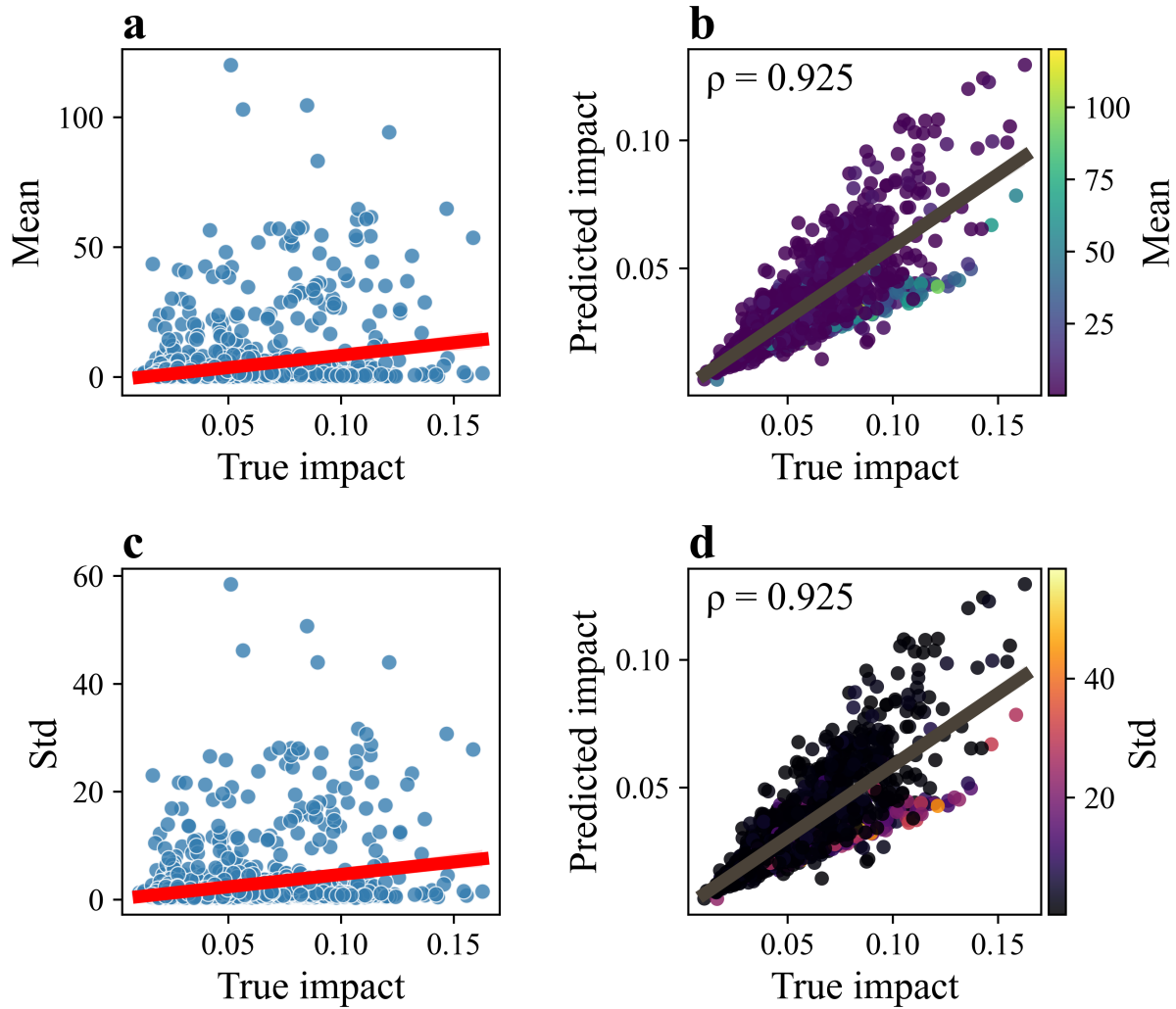

**Figure S4: Correlation between ground truth gene KO impact and gene-level statistics in the K562 cell type.** **(a)** True KO impact versus mean gene count across cells. **(b)** Predicted versus true KO impact, colored by mean gene count across cells. **(c)** True KO impact versus the across-cell standard deviation (Std) of gene count. **(d)** Predicted versus true KO impact, colored by the across-cell Std. Genes with both high mean counts and high standard deviation of counts tend to lie below the fitted line, indicating these genes are challenging to predict, and the model underestimate their KO impact.

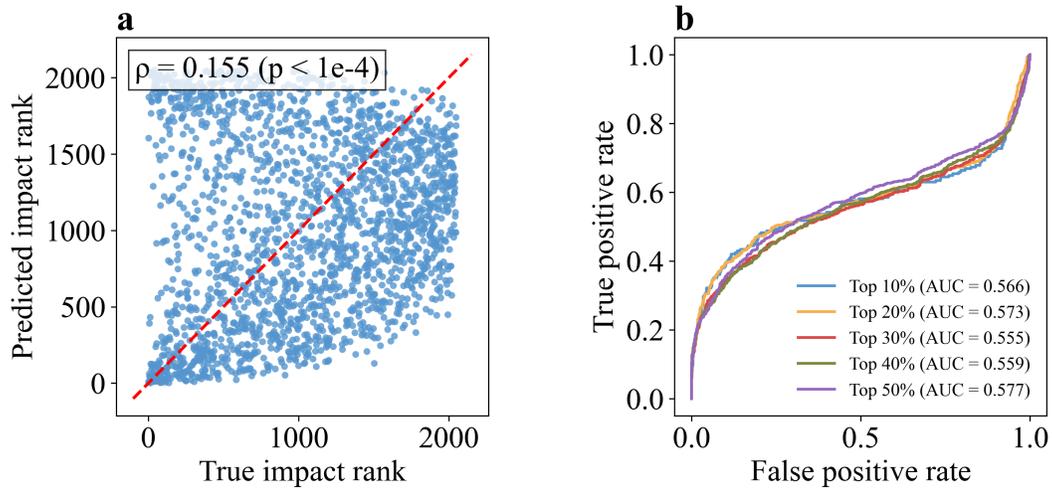

**Figure S5: Cross-cell type evaluation of the model using Perturb-seq data.** Training on the RPE1 cell type and predicting KO impacts in the K562 cell type resulted in poor performance.